\documentclass[referee]{raa}
\usepackage{graphicx,times}
\usepackage{natbib}
\usepackage{amssymb,amsmath}
\bibpunct{(}{)}{;}{a}{}{,}
\makeatletter

\newcommand{\Rmnum}[1]{\expandafter\@slowromancap\romannumeral #1@}

\makeatother
\def\degree{{}^{\circ}}

\usepackage[pagebackref=true]{hyperref}
\hypersetup{pdftitle = The title of my PDF, pdfauthor = My name, pdfsubject= The subject, pdfkeywords = keyword1 keyword2 keyword3}
\hypersetup{colorlinks = true, linkcolor = green, anchorcolor = red, citecolor = blue, filecolor = red, pagecolor = red, urlcolor = red}

\begin{document}
\title{Are Pulsar Giant Pulses Induced by Re-emission of Cyclotron Resonance Absorption?}
\volnopage{ {\bf 2015} Vol.\ {\bf X} No. {\bf XX}, 000--000}
\setcounter{page}{1}

\author{Ji-Guang Lu\inst{1,2}, Wei-Yang Wang\inst{3,4}, Bo Peng\inst{1,2}, Ren-Xin Xu\inst{5,6}}
\institute{CAS Key Laboratory of FAST, National Astronomical Observatories, Chinese Academy of Sciences, Beijing 100101, China; {\it lujig@nao.cas.cn} \\
\and
Guizhou Radio Astronomy Observatory, Chinese Academy of Sciences, Guiyang 550025, China; \\
\and
Key Laboratory for Computational Astrophysics, National Astronomical Observatories, Chinese Academy of Sciences, 20A Datun Road, Beijing 100101, China; \\
\and
University of Chinese Academy of Sciences, Beijing 100049, China\\
\and
Department of Astronomy, Peking University, Beijing 100871, China; \\
\and
Kavli Institute for Astronomy and Astrophysics}

\abstract{%
It is conjectured that coherent re-emission of cyclotron resonance
absorption could result in pulsar giant pulses.
This conjecture seems reasonable as it can naturally explain
the distribution of pulsars with giant pulses on the $P$-$\dot{P}$ diagram.
\keywords{pulsars: general --- radiation mechanisms: non-thermal --- plasmas --- magnetic fields}
}
\authorrunning{Lu, Wang, Peng and Xu}
\titlerunning{Cyclotron Resonance Absorption and Giant Pulse}
\maketitle

\section{Introduction}
\label{sect:intro}

Pulsars were discovered more than half a century ago, but their radiation mechanism,
in both radio and X/$\gamma-$ray bands,  remains uncertain.
The coherent radio radiation of a pulsar is generated by charged particles
(i.e., dense plasma) in its magnetosphere at low altitude (e.g., \citealt{Luo95,Gil04}).
While the radio wave propagates in the magnetosphere, it interacts with
the less-dense plasma.
Two eigen polarization modes of the radiation exist in this magnetized low-density
plasma~\citep{wang10}, and one of them at a certain frequency would be
absorbed by the plasma via the cyclotron resonance effect~\citep{rafi05}.


We conjecture that coherent re-emission in the cyclotron resonance region could
result in giant pulses, leading to pulsars with giant pulses being located
in certain regions of the the $P$-${\dot P}$ diagram.
%
%
This paper is organized as follows: cyclotron resonance absorption is introduced in
Section~\ref{sect:cyclo}, and discussion and conclusion are presented in Section~\ref{sect:disc}
and Section~\ref{sect:sum}.

\section{Cyclotron Resonance Absorption}
\label{sect:cyclo}

Considering a plasma with magnetic field $\mathbf{B}=(0, B_y, B_z)$ in rectangular coordinates,
radiation with wave vector $\mathbf{k}$
propagating along the \textit{z}-axis ($\hat{\mathbf{k}}=\hat{\mathbf{e}}_z$)
has two eigenmodes,
\begin{equation}
k_{\mathrm{R(L)}}=\frac{\omega}{c}\left[1-\frac{\omega_{\mathrm{p}}^2}{\omega^2}\frac{1}
{1-\frac{1}{2}\frac{\omega_{y}^2}{\omega^2-\omega_{\mathrm{p}}^2}\mp
\sqrt{\frac{\omega_{z}^2}{\omega^2}+\left(\frac{1}{2}\frac{\omega_{y}^2}
{\omega^2-\omega_{\mathrm{p}}^2}\right)^2}}\right]^{\frac{1}{2}},
\label{wv}
\end{equation}
$$
\omega_{y}=\frac{B_y}{B}\omega_B,~~~
\omega_{z}=\frac{B_z}{B}\omega_B,
$$
where $\omega$ is the circular frequency of the radiation,
$\omega_{\mathrm{p}}=\sqrt{\frac{4\pi n_{\mathrm{p}}q_{\mathrm{p}}^2}{m_{\mathrm{p}}}}$
is the plasma frequency,
$\omega_{B}=\frac{Bq_{\mathrm{p}}}{m_{\mathrm{p}}c}$ is the cyclotron frequency,
$B=\left|\mathbf{B}\right|=\sqrt{B_y^2+B_z^2}$ are the strength of the magnetic field,
$m_{\mathrm{p}}$ and $q_{\mathrm{p}}$ are the mass and the charge of the particles in plasma,
$n_{\mathrm{p}}$ is the number density of the particles,
and $c$ is the speed of light.
In fact, either of the two eigenmodes is an elliptical polarization mode,
and the subscript of the wave vector R(L) refers to the wave, being a right (left)
circularly polarized wave for positive $B_z$.
The electric field of the wave $\mathbf{E}=(E_x, E_y, E_z)$ for the two
eigenmodes follows
\begin{equation}
E_{y,~\mathrm{R(L)}}=\mathrm{i}\frac{\omega_{z}}{\omega}\frac{1}
{1-\frac{1}{2}\frac{\omega_{y}^2}{\omega^2-\omega_{\mathrm{pe}}^2}\pm
\sqrt{\frac{\omega_{z}^2}{\omega^2}+\left(\frac{1}{2}\frac{\omega_{y}^2}
{\omega^2-\omega_{\mathrm{pe}}^2}\right)^2}}E_{x,~\mathrm{R(L)}}.
\label{we}
\end{equation}
The frequency of observed radiation is always higher than the local plasma frequency
in the magnetosphere of the pulsar, which decreases along the radiation propagating path.
With Eq.~\ref{wv}, the L-mode wave can always freely propagate, while the
wave vector of the R-mode wave may take on an imaginary value for some magnetic fields,
where the radiation frequency is close to the cyclotron frequency.
In fact, the local plasma frequency is very low during the absorption process,
the two eigenmodes change little with Eq.~\ref{we}.
Then the absorbed component remains unchanged.
With the above assumptions, the optical depth $\tau$ of the R-mode wave can be calculated,
\begin{equation}
\tau\approx-\pi^2n_{\mathrm{p}}q_{\mathrm{p}}\frac{1+\frac{2\omega_{cz}^2}{\omega_{cy}^2}}
{\left(1+\frac{\omega_{cz}^2}{\omega_{cy}^2}\right)^{\frac{3}{2}}}
\frac{1}{\frac{\mathrm{d}B_y}{\mathrm{d}z}}.
\label{od}
\end{equation}
In fact, the term $\left(1+\frac{2\omega_{cz}^2}{\omega_{cy}^2}\right)/
\left(1+\frac{\omega_{cz}^2}{\omega_{cy}^2}\right)^{\frac{3}{2}}$ is
approximately 1 for $\frac{B_z}{B_y}<3.73$, i.e., the induced angle between $<\hat{\mathbf{e}}_z$
and $\hat{\mathbf{B}}>$ is less than 75$\degree$, which can be always satisfied at the absorbing location.
Considering a dipole field $B_y\sim B\sim\frac{\mu}{r^3}$ (where
$r$ is the distance between the pulsar center and the absorption point,
$\mu$ is the magnetic dipole moment of the pulsar), a particle charge density
$n_{\mathrm{p}}q_{\mathrm{p}}=\lambda\rho_{\mathrm{GJ}}=\lambda\frac{B}{Pc}$
(where $\lambda$ is the ratio of the particle charge density and the GJ charge
density given in~\citealt{gold69}, $P$ is the period of the pulsar),
Eq~\ref{od} can be expressed as below,
$$
\tau\sim\frac{\pi^2\lambda}{3cP}\left(\frac{q_{\mathrm{p}}\mu}
{\omega m_{\mathrm{p}}c}\right)^{\frac{1}{3}}
$$
\begin{equation}
\approx0.155\lambda\times\left(\frac{P}{1\,\mathrm{s}}\right)^{-\frac{5}{6}}
\left(\frac{\dot{P}}{10^{-14}\,\mathrm{s\cdot s^{-1}}}\right)^{\frac{1}{6}}
\left(\frac{\nu}{1\,\mathrm{GHz}}\right)^{-\frac{1}{3}}~~~~~~~\mathrm{(for~electrons),}
\end{equation}
\begin{equation}
\approx0.0127\lambda\times\left(\frac{P}{1\,\mathrm{s}}\right)^{-\frac{5}{6}}
\left(\frac{\dot{P}}{10^{-14}\,\mathrm{s\cdot s^{-1}}}\right)^{\frac{1}{6}}
\left(\frac{\nu}{1\,\mathrm{GHz}}\right)^{-\frac{1}{3}}~~~~~~~\mathrm{(for~protons).}
\end{equation}
The parameter $\lambda$ can be either larger than 1
(e.g., the case of incomplete charge separation, or the case of
outward particle flow)
or less (e.g., in the gap).

\begin{figure}
\centering
\includegraphics[width=0.8\textwidth,angle=0]{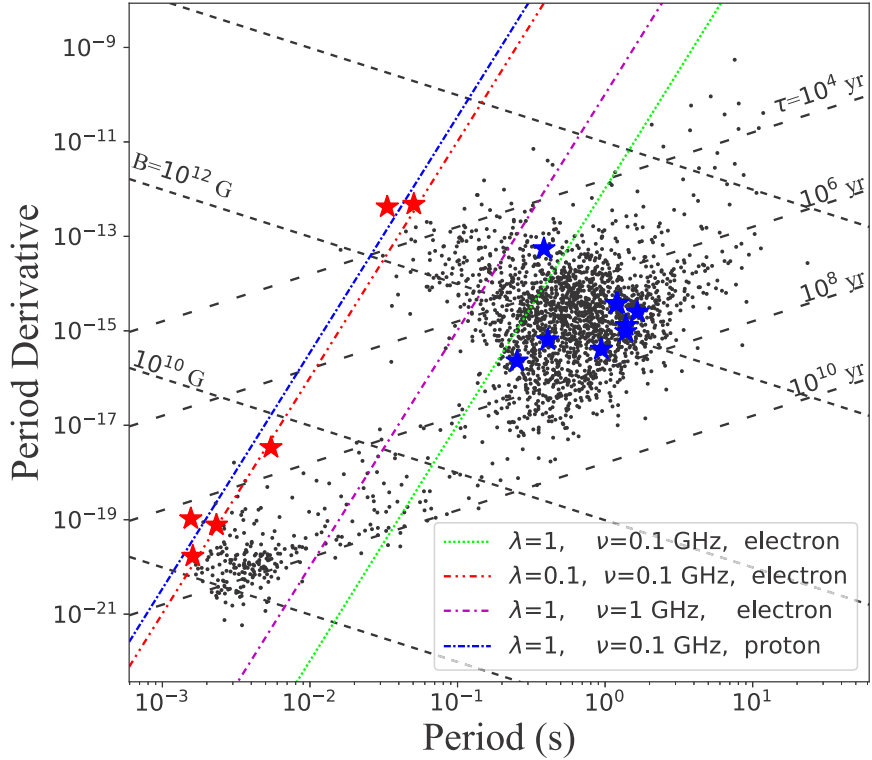}
\caption{The pulsars with giant pulses are marked on the $P$-$\dot{P}$ diagram,
where the red and blue stars represent the pulsars with giant pulses and giant micropulses, respectively.
The dotted, dash-dotted and dash-dot-dotted lines mark the boundaries $\tau=1$ for different parameters,
where the red line is the reference calculated for electrons with $\lambda=1$, $\nu=0.1$\,GHz,
and the blue, magenta and green lines change with the $\lambda$, $\nu$ and particles respectively.
}
\label{f1}
\end{figure}

If the optical depth is larger than 1, the radiation will be absorbed significantly.
Thus a boundary can be defined according to the optical depth of 1.
Some boundaries are drawn on the $P$-$\dot{P}$ diagram in Figure.~\ref{f1}.
The red stars represent the pulsars with giant pulses,
whereas the blue stars represent those with giant micropulses
(\citealt{cair04}, they are also called type II giant pulses in \citealt{wang19}).
In the figure, lines with different parameters, e.g. radiation frequency $\omega$,
ratio factor $\lambda$, for electrons and protons are drawn.
Coincidentally, some of these boundaries can be used to distinguish
pulsars with observable giant pulses from other pulsars.
Therefore, it can be speculated that the cyclotron resonance absorption
may re-emit the giant pulse.
As demonstrated in Figure.~\ref{f1}, both the electrons and protons can
be the proper absorber.

\section{Discussion}
\label{sect:disc}

The absorbed energy can be converted into coherent radiation and re-emitted.
The radiation mechanism may be cyclotron resonance~\citep{wang19} or
cyclotron-Cherenkov resonance~\citep{lyut99}, or some other process.
If the radiation turns into the observed giant pulse,
the time scale of the emission should be limited.
\cite{hank03} found that the giant pulse of the Crab pulsar has nanosecond structure,
which means that the radiation time scale should be at the nanosecond level.
It requires a high radiant power.
As the radiation power of a giant pulse is extraordinarily stronger than that of the normal pulses,
the emission rate of giant pulses must be larger than the absorption rate of cyclotron resonance.
Therefore, the radiation of a giant pulse should be energy released
after its accumulation via cyclotron resonance absorption over some period.
It should be an unstable process which may be related to the phenomenon
of self organized criticality, which can explain the power-law flux distribution
of giant pulses~\citep{popo07}.

It is known that the giant pulses in some pulsars are in phase or slightly offset in
phase with the high energy non-thermal radiation, such as the Crab pulsar (PSR B0531$+$21,
~\citealt{abdo10}), PSR B1937$+$21~\citep{cusu03}; for some pulsars, only part
of the giant pulses are in phase with the high energy non-thermal radiation, such as PSR B1957$+$20~\citep{main17,guil12}; for the other pulsars, the peak position at high energy band and the giant radio pulses are separated, such as PSR B1821$-$24A~\citep{john13} and  PSR B0540$-$69~\citep{john04}.
On the other hand, the giant pulses from Crab pulsar has no significant correlation with the X-ray radiation~\citep{hito18}, it means that the giant pulses and X-ray radiation of Crab pulsar may have different spatial origination.
It should be noted that the pulse phase are always related to the radiation altitude in the dipole field, therefore the numerous observations implicate that the giant pulse and high energy non-thermal radiation may have similar radiation altitude.
The X-ray radiation of Crab pulsar can be reconstructed in both slot gap~\citep{hard08} and annular gap model~\citep{du12}, and radiation altitudes are both high.
In fact, the cyclotron resonance occurs when the cyclotron frequency is similar to the radiation frequency and always takes place in the high altitude region.
Thus the giant pulse can have similar radiation phase to the high energy radiation.

\cite{hank16} found zebra-pattern-like spectral bands in
giant pulses of inter pulses from the Crab pulsar.
It is worth noting that those spectral bands were only observed above 5\,GHz,
and all the giant pulses observed above 5\,GHz exhibited the zebra-pattern-like bands.
This critical frequency implies that there is a physical limit,
and the frequency of cyclotron absorption may be one answer.
Based on the lines in Figure~\ref{f1}, this speculation prefers the absorber
to be the electron, or to be the protons when the $\lambda$ parameter is bigger than 1 in the large outwards particle flow.

\cite{wang19} pointed out that the pulsars with giant pulses have a similar
magnetic field at the light cylinder, which implies that the giant pulses may be
generated near the light cylinder.
\cite{lyub19} and \cite{phil19} also build models that the magnetic reconnection near light cylinder can lead to radio nano-shots.
In fact, the Eq.~\ref{od} can be rewritten as the following,
\begin{equation}
\tau\approx\frac{\pi\lambda}{6}\left(\frac{q_{\mathrm{p}}}
{\omega m_{\mathrm{p}}c}\right)^{\frac{1}{3}}B_{\mathrm{LC}}^{\frac{1}{3}},
\end{equation}
where $B_{\mathrm{LC}}=\left(\frac{2\pi}{cP}\right)^3\mu$ is close to
the magnetic field strength at the light cylinder.
This means that the boundary for the optical depth of the R-mode wave
can be expressed in terms of the magnetic field at the light cylinder.
So the giant pulses can also be generated from the inner region instead of near the light cylinder based on the locations of pulsars with giant pulses in the $P$-$\dot{P}$ diagram.

The magnetic field of any pulsar is always twisted due to the pulsar rotation.
Then the drift velocity of the particles in the magnetosphere which runs parallel
to the local Poynting vector must have a radial component.
The Poynting vector of a dipole rotator generally points outwards, thus the
particles in the magnetosphere tend to move outwards.
As magnetic field distortion is more significant in the outer location,
the loss of particles becomes more serious.
Therefore, the particle charge density may deviate more from the GJ charge
density by being farther from the pulsar, and $\lambda$ becomes smaller.
For the electron absorption, the magnetic field is $\sim10^3$\,G, which is
almost the value at the light cylinder, indicating that the $\lambda$ is smaller
than 1.
Whereas, for the proton absorption, the magnetic field is $\sim10^6$\,G, and
the parameter $\lambda$ should be bigger.

With Eq.~\ref{we}, the polarization degree of each mode can be calculated.
It is found that the circular polarization degree of each mode in the absorption
region is very high.
If the R-mode wave is absorbed, the left radiation containing only the
L-mode component should be highly circularly polarized.
However, the circular polarization of radiation from some pulsars
with giant pulses is relatively low, e.g. the radiation from the Crab pulsar~\citep{moff99}
and PSR B1937$+$21~\citep{dai15}.
This may result from the difference between the core (the region
surrounded by the critical field line) and the annular region (the region
between the critical field line and the last opening field line).
The annular region may be free of plasma, and allow radiation
to propagate freely~\citep{du10,du12},
while the core region may be filled by charged particles.
The radiation of normal pulses can be generated from the annular region,
while giant pulses come from the core region.

\section{Summary}
\label{sect:sum}

To summarize, cyclotron resonance absorption is a possible power
source for the giant pulse, that might naturally explain the distribution
of pulsars with giant pulses on the $P$-$\dot{P}$ diagram.
The radiation mechanism producing giant pulses is not discussed in detail in this article, and none of the properties of giant pulses (short time duration and high power, nano-shot structure, polarization) are addressed in this model;
it will be studied in the future.

In this paper, giant pulses of radio pulsars are investigated from a theoretical point
of view, which could be applicable to future observational investigations.
It is worth noting that, with superior sensitivity compared to any other
single-dish radio telescope, Chinese FAST may help in the study of giant pulses.
We would then anticipate a FAST era of pulsar science to come~\citep{peng00a,peng00b,Jiang19,wanghg19,Lu20}.
%

\begin{acknowledgements}
This work is supported by the National Key R\&D Program of China under grant number 2018YFA0404703,  NSFChina (Grant Nos. 11673002 and U1531243), and the Open Project Program of the Key Laboratory of FAST, NAOC, Chinese Academy of Sciences.
J.G. Lu acknowledges the support of the FAST FELLOWSHIP from Special Funding budgeted and administrated by the Center for Astronomical Mega Science, Chinese Academy of Sciences (CAMS).
B. Peng acknowledge the CAS-MPG LEGACY funding ``Low-Frequency Gravitational Wave Astronomy and Gravitational Physics in Space''.
Thanks Richard Strom for giving suggestions on the revision.
\end{acknowledgements}


\begin{thebibliography}{10}


\bibitem[Abdo(2010)]{abdo10}
Abdo, A. A., Ackermann, M., Ajello, M., et al. 2010, ApJ, 708, 1254

\bibitem[Cairns(2004)]{cair04}
Cairns, I.~H.\ 2004, \apj, 610, 948

\bibitem[Cusumano et al.(2003)]{cusu03}
Cusumano, G., Hermsen, W., Kramer, M., et al.\ 2003, \aap, 410, L9

\bibitem[Dai et al.(2015)]{dai15}
Dai, S., Hobbs, G., Manchester, R.~N., et al.\ 2015, \mnras, 449, 3223

\bibitem[Du et al.(2010)]{du10}
Du, Y.~J., Qiao, G.~J., Han, J.~L., et al.\ 2010, \mnras, 406, 2671

\bibitem[Du et al.(2012)]{du12}
Du, Y.~J., Qiao, G.~J., \& Wang, W.\ 2012, \apj, 748, 84

\bibitem[Gil et al.(2004)]{Gil04}
Gil, J., Lyubarsky, Y., \& Melikidze, G.~I.\ 2004, \apj, 600, 872

\bibitem[Goldreich and Julian(1969)]{gold69}
Goldreich, P. \& Julian, W. H.\ 1969, \apj, 157, 869

\bibitem[Guillemot et al.(2012)]{guil12}
Guillemot, L., Johnson, T.~J., Venter, C., et al.\ 2012, \apj, 744, 33

\bibitem[Hankins et al.(2003)]{hank03}
Hankins, T.~H., Kern, J.~S., Weatherall, J.~C., et al.\ 2003, \nat, 422, 141

\bibitem[Hankins et al.(2016)]{hank16}
Hankins, T.~H., Eilek, J.~A., \& Jones, G.\ 2016, \apj, 833, 47

\bibitem[Harding et al.(2008)]{hard08}
Harding, A.~K., Stern, J.~V., Dyks, J., et al.\ 2008, \apj, 680, 1378

\bibitem[Hitomi Collaboration et al.(2018)]{hito18}
Hitomi Collaboration, Aharonian, F., Akamatsu, H., et al.\ 2018, \pasj, 70, 15

\bibitem[Jiang et al.(2019)]{Jiang19}
Jiang, P., Yue, Y. L., Gan, H. Q., et al\ 2019, Sci. China-Phys. Mech. Astron., 62, 959502

\bibitem[Johnston et al.(2004)]{john04}
Johnston, S., Romani, R.~W., Marshall, F.~E., et al.\ 2004, \mnras, 355, 31

\bibitem[Johnson et al.(2013)]{john13}
Johnson, T.~J., Guillemot, L., Kerr, M., et al.\ 2013, \apj, 778, 106

\bibitem[Lu et al.(2020)]{Lu20}
Lu, J. G., Li, K. J. and Xu, R. X.\ 2020, Sci. China-Phys. Mech. Astron., 63, 229531

\bibitem[Luo \& Melrose(1995)]{Luo95}
Luo, Q., \& Melrose, D.~B.\ 1995, \mnras, 276, 372

\bibitem[Lyubarsky(2019)]{lyub19}
Lyubarsky, Y.\ 2019, \mnras, 483, 1731

\bibitem[Lyutikov et al.(1999)]{lyut99}
Lyutikov, M., Blandford, R.~D., \& Machabeli, G.\ 1999, \mnras, 305, 338

\bibitem[Main et al.(2017)]{main17}
Main, R., van Kerkwijk, M., Pen, U.-L., et al.\ 2017, \apjl, 840, L15

\bibitem[Moffett \& Hankins(1999)]{moff99}
Moffett, D.~A. \& Hankins, T.~H., 1999, \apj, 522, 1046

\bibitem[Peng et al.(2000a)]{peng00a}
Peng, B., Strom, R.~G., Nan, R. et al.\ 2000, Perspectives on Radio Astronomy: Science with Large Antenna Arrays, 25

\bibitem[Peng et al.(2000b)]{peng00b}
Peng, B., Nan, R., \& Su, Y.\ 2000, \procspie, 45

\bibitem[Philippov et al.(2019)]{phil19}
Philippov, A., Uzdensky, D.~A., Spitkovsky, A., et al.\ 2019, \apjl, 876, L6

\bibitem[Popov \& Stappers(2007)]{popo07}
Popov, M.~V., \& Stappers, B.\ 2007, \aap, 470, 1003

\bibitem[Rafikov \& Goldreich(2005)]{rafi05}
Rafikov, R.~R., \& Goldreich, P.\ 2005, \apj, 631, 488

\bibitem[Wang et al.(2010)]{wang10}
Wang, C., Lai, D., \& Han, J.\ 2010, \mnras, 403, 569

\bibitem[Wang, Qiao, et al.(2019)]{wanghg19}
Wang, H., Qiao, G. J., Du, Y. J., et al.\ 2019, RAA, 19, 21

\bibitem[Wang et al.(2019)]{wang19}
Wang, W., Lu, J., Zhang, S., et al.\ 2019, Sci. China-Phys. Mech. Astron., 62, 979511

\end{thebibliography}
\end{document}